# DEPENDENCE OF THE SOURCE DEACTIVATION FACTOR ON THE EARTHQUAKE MAGNITUDE

A.V. Guglielmi, O.D. Zotov

*Schmidt Institute of Physics of the Earth, Russian Academy of Sciences, Moscow, Russia,*
*guglielmi@mail.ru, ozotov@.inbop.ru*

**Abstract**

Three fundamental laws of the physics of earthquakes, bearing the names of their discoverers Omori, Gutenberg, Richter and Bath, are widely used in original, review, monographic, and encyclopedic literature. In this paper, we have tried to look at laws from an unusual point of view. The idea behind our approach is as follows. Each of the three laws was established in the course of work on solving a specific research problem, and the result was expressed in a specific mathematical form. We asked ourselves whether it is possible to modify the statement of the problem, and/or modify the form of expression of the law in order to see new facets of a well-known scientific statement? The result was successful in relation to Omori's law, according to which the frequency of aftershocks hyperbolically decreases over time. The modified formulation of the problem was to find a differential equation describing the time evolution of aftershocks. We found a simple differential model, within the framework of which the concept of deactivation of the earthquake source is introduced. The question of dependence of the deactivation factor on the magnitude of main shock is analyzed theoretically and experimentally. A monotonic decrease in the deactivation factor with an increase in the magnitude of the main shock has been reliably established. Other possible applications of the proposed methodological technique are considered. In particular, it is indicated how to overcome the contradiction between the Omori law and the law of energy conservation.

## 1. Introduction

Three empirical laws are widely used in the physics of earthquakes, namely, Omori law [1], Gutenberg-Richter law [2], and Bath law [3]. The richest encyclopedic, monographic and review literature is devoted to these laws (e.g., see [4–8]). In many original scientific papers, three laws are used in the systematization of observations, in the search for options for the theoretical interpretation of laws, as well as in the search for new laws, patterns and properties of earthquakes – these extremely complex and dangerous natural phenomenon.

In this paper, we want to look at laws from an unusual point of view. It is quite clear that each law was discovered in the process of work, in which a specific research problem was posed, and the law itself was formulated in certain physical and mathematical terms [1–3]. We tried, in pursuit of purely methodological goals, to vary the setting of the problem and/or reformulate the law. Sometimes attempts of this kind give interesting results.

Our approach has been particularly successful with regard to Omori law. Omori was faced with the task of finding a mathematical expression describing the decrease in the frequency $n(t)$ of aftershocks over time. He very successfully chose the algebraic formula $n = k/t$, in which $k$ is some phenomenological parameter, and $t \geq t_0$, where $t_0 > 0$ is a more or less arbitrary moment of the beginning of the calculation of the frequency of aftershocks. We have reformulated the problem: find a differential equation describing the time evolution of aftershocks [9]. It turned out that Omori law is obtained from the solution of equation

$$\frac{dn}{dt} + \sigma n^2 = 0 . \tag{1}$$

Here $\sigma$ is the so-called deactivation factor of the earthquake source, "cooling down" after the main shock. The concept of deactivation of the source and the writing of Omori law in the form of differential equation (1) opens up interesting possibilities for searching for generalizations of the theory and for searching for new ways of analyzing experimental data (see, for example, [10–17]).

In this work, we focused on the possible dependence of deactivation factor $\sigma$ on the magnitude of main shock $M_0$. In the theoretical analysis, we used not only the Gutenberg-Richter law for aftershocks, but also the dependence of the number of aftershocks on the magnitude of main shock. The experimental study fully confirmed the theoretical prediction.

In the course of our study, the question arose about the compatibility of the aftershocks equation (1) with the energy conservation law. In this regard we, firstly, pointed out possible generalizations of equation (1) and, secondly, discussed the Bath law, which, apparently, needs additional research.

## 2. Theory

Let us change the variable $n \to g = 1/n$ in equation (1) and rewrite the Omori law in its simplest form

$$\frac{dg}{dt} = \sigma. \tag{2}$$

We use the solution $g = g_0 + \sigma t$ of equation (2) to calculate the number of aftershocks $N$ on the interval $[0, T]$:

$$N = \int_0^T \frac{dt}{g(t)} = \frac{1}{\sigma} \ln\left[1 + \left(\frac{T}{g_0}\right)\sigma\right]. \tag{3}$$

We see that $N$ is a monotonically decreasing function of $\sigma$. In other words,

$$\frac{dN}{d\sigma} < 0. \tag{4}$$

On the other hand, experience shows that, on average, the number of aftershocks $N$ is higher, the higher the magnitude of the main shock $M_0$:

$$\frac{dN}{dM_0} > 0. \tag{5}$$

Inequalities (4) and (5) imply the inequality

$$\frac{d\sigma}{dM_0} < 0. \tag{6}$$

So, the theory predicts that the source deactivation coefficient $\sigma$ is a monotonically decreasing function of the magnitude of the main shock $M_0$.

## 3. Experiment

To experimentally verify inequality (6), we used data on earthquakes that occurred on Earth from 1973 to 2019 and were registered in the world catalog of earthquakes USGS/NEIC (https://earthquake.usgs.gov). There were found 2508 main shocks with a magnitude of $M_0 \geq 6$ and a hypocenter depth not exceeding 250 km. For each main shock, a circular epicentral zone was determined by the formula

$$\lg L = 0.43 M_0 - 1 .. \qquad (7)$$

Here, the radius of the zone $L$ is expressed in kilometers [18]. According to our definition, the event is an earthquake with a magnitude $M < M_0$, which occurred in the epicentral zone in the interval of $\pm 20$ hours relative to the moment of the main shock. A total of 1618 events were accumulated before, and 30837 events after the main shock.

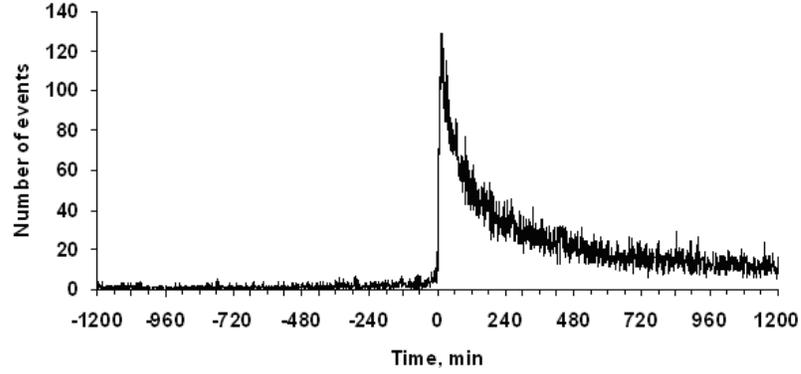

**Fig. 1.** Distribution of events in time. The graph is constructed using the epoch superposition method. Time is counted from the moment of the main shock.

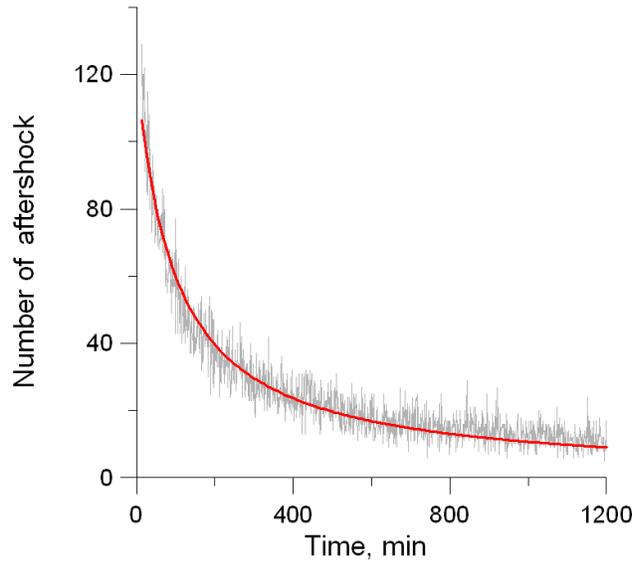

**Fig. 2.** The distribution of aftershocks in time. The red curve shows Omori hyperbolic law.

Figure 1 gives an idea of the distribution of events over time. A similar distribution for aftershocks is shown in Figure 2. The red curve is drawn in accordance with the Omori law $n = k/t$, $t \geq t_0$. The curve parameters are as follows: $k = 11760$, $t_0 = 100$ min. The coefficient of determination is very high: $R^2 = 0.93$.

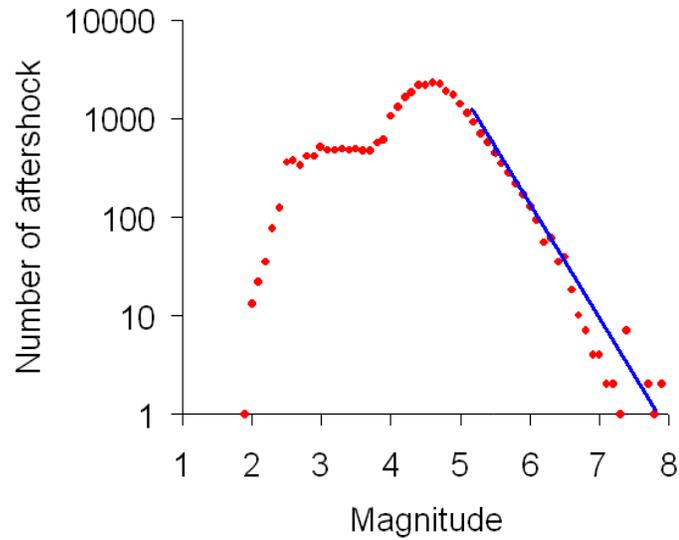

**Fig. 3.** Distribution of aftershocks by magnitude. The Gutenberg-Richter distribution is plotted by straight line.

The distribution of aftershocks by magnitude is shown in Figure 3. The straight line in the figure approximates the representative part of the distribution that satisfies the Gutenbern-Richter law

$$\lg N = a - bM. \tag{8}$$

Here $a = 9.14$, $b = 1.18$, and $R^2 = 0.99$.

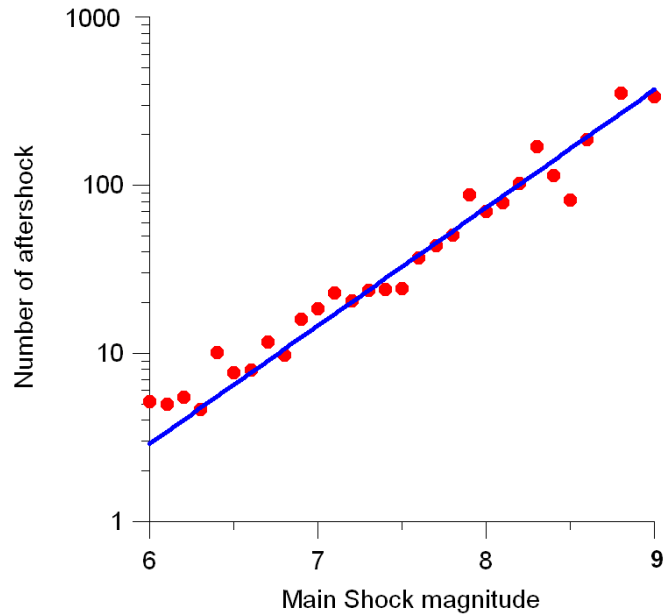

**Fig. 4.** Dependence of the number of aftershocks on the magnitude of the main shock in the interval from 0 to 20 hours.

We checked whether inequality (5) holds. The result is shown in Figure 4. The dependence $N(\mathrm{M}_0)$ looks like "mirror" in relation to the Gutenbeog-Richter law $N(\mathrm{M})$. (It remotely resembles a mirror reprise, for example, in Les Préludes.) The formula

$$\lg N = -\alpha + \beta\, \mathrm{M}_0 \qquad (9)$$

approximates the found dependence. Here $\alpha = 3.7$, $\beta = 0.7$. The coefficient of determination is very high ($R^2 = 0.92$), so inequality (5), which follows from (9), can be considered quite reasonable.

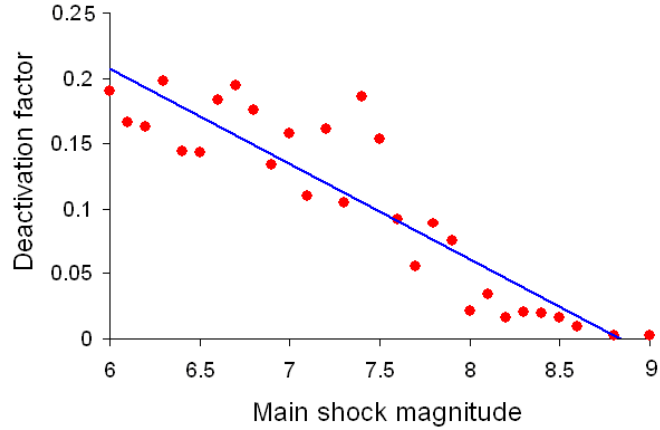

**Fig. 5.** Dependence of the deactivation factor of the earthquake source on the magnitude of the main shock.

Figure 5 shows the result of measurements of $\sigma$ at different values of $\mathrm{M}_0$. To measure the deactivation factor, we used the technique developed during the compilation of the Atlas of Aftershocks [12, 13]. We see that, on average, $\sigma$ decreases monotonically with the increase in $\mathrm{M}_0$. The dependence $\sigma(\mathrm{M}_0)$ is approximated by the formula

$$\sigma = A - B\mathrm{M}_0, \qquad (10)$$

where $A = 0.64$, $B = 0.07$ with a sufficiently high coefficient of determination $R^2 = 0.82$. Thus, the inequality $d\sigma/d\mathrm{M}_0 < 0$ is reliably confirmed by direct measurements.

## 4. Contradiction

According to Bath law, the magnitudes of aftershocks of a particular event forms a set $\{\mathrm{M}\}$ of numbers not exceeding the value of $\mathrm{M}_{\max} = \mathrm{M}_0 - \Delta \mathrm{M}$. In the literature, the value of $\Delta \mathrm{M} = 1.1 - 1.2$ is usually indicated (see, for example, [3, 4, 19]). We were wondering what can be said about the *minimum* magnitude in the $\{\mathrm{M}\}$ set, and how does it relate to the magnitude of the main shock? We don't know the answer. And the question itself may seem far-fetched. But, strictly speaking, if a finite minimum magnitude exists, then from a formal point of view, the Omori law [1] contradicts the law of energy conservation. Indeed, if $W$ is the power of the aftershock flow and the Omori law $n = k/t$ is fulfilled, then the total energy of the aftershocks

$$E = \lim \int_{t_0}^{t} W(t') dt' \propto \lim \ln t^k, \quad t \to \infty \qquad (11)$$

is equal to infinity.

The formulation of law in the form of the evolution equation (1) avoids a contradiction. Indeed, it suffices to assume that the deactivation factor $\sigma(t)$ is time dependent. Then, instead of the Omori formula, we will have

$$n(t) = \frac{n_0}{1 + n_0 \tau(t)}, \qquad (12)$$

where $n_0 = n(0)$ and

$$\tau = \int_0^t \sigma(t') dt. \qquad (13)$$

In this case, the energy is finite if the deactivation factor increases asymptotically with time rather quickly.

So we assume that $\sigma$ is time dependent. Does this not contradict the linear dependence $g(t) = g_0 + \sigma t$, which we used in the previous two sections of the paper, and which is true only if $\sigma = \text{const}$? The point is as follows. The experience of compiling the Atlas of Aftershocks [12, 13, 17] showed that the deactivation coefficient really undergoes complex variations during the relaxation of the source after the formation of a main rupture. However, at the first stage of evolution $\sigma = \text{const}$. The time interval at which $\sigma = \text{const}$ we call the Omori epoch. The duration of the Omori epoch varies from several days to many tens of days. The $\sigma$ measurements we used to construct Figure 5 are quite correct in this respect, since they were carried out in the interval from 0 to 15 hours after the main shock, i.e. in the Omori era.

Note that in [8] the tendency towards an increase in the duration of the Omori epoch with the increase in the magnitude of the main shock is indicated. In our opinion, this trend deserves further study.

## 5. Discussion

Writing Omori law in the form of equation (1) opens up interesting possibilities for generalizations of the theory. The natural generalization is the logistic equation

$$\frac{dn}{dt} = n(\gamma - \sigma n). \qquad (14)$$

Instead of (14), one can use the equation

$$\frac{dg}{dt} + \gamma g = \sigma. \qquad (15)$$

Here $\gamma$ is the second phenomenological parameter of our theory. Concerning the earthquake modeling based on the logistic equation see [15, 17, 20].

We find a broader generalization when searching for models of the space-time distribution of aftershocks $n(x,t)$. An interesting result was obtained by adding the term $\kappa n''$ to the right side of equation (14):

$$\frac{\partial n}{\partial t} = n(\gamma - \sigma n) + \kappa \frac{\partial^2 n}{\partial x^2}. \qquad (16)$$

Equation (16) has solutions in the form of slow nonlinear waves [21], which resemble the wavelike structure of the spatio-temporal distribution of aftershocks found in [11]. The three-parameter model (16) is sometimes called $\gamma\sigma\kappa$-model.

In connection with the Omori law, it is worth mentioning work [22]. In this work, the authors drew attention to the analogy between the Barkhausen effect in ferroelectric and such a phenomenon as relaxation of the earthquake source due to repackaging of structural elements of the lithosphere and crackling noise in the form of aftershocks accompanying relaxation.

Finally, speaking about the Omori law $n = k/t$ [1], one cannot fail to mention the two-parameter Hirano-Utsu law $n = k/t^p$ [23, 24]. Numerous measurements indicate that the parameter $p$ varies from case to case and from place to place from 0.7 to 1.5, with $p = 1.1$ on average (see review [7] and the literature cited therein). Let's change the variable $n \to g = 1/n$. The Hirano-Utsu law will appear in the form $g = t^p/k$. The deactivation factor $\sigma = dg/dt$ in this case is $\sigma = (p/k)t^{p-1}$. Consequently, for $p \neq 1$, the Omori epoch does not exist, which sharply contradicts the observations [12, 13]. Thus, the Hirano-Utsu law is unacceptable. In the Omori epoch, the Omori law is fulfilled in the original formulation [1], or in the form of the evolution equation (1) with $\sigma = \text{const}$. The consideration given here inclines us to an affirmative answer to the following question:

Isn't the Omori formula an expression of the fundamental law of earthquake physics, while the Hirano-Utsu formula is just a fitting formula?

Let's move on to discussing the Bath law. Let us point out here to the substantive works [19, 25, 26] devoted to experimental verification and the search for a theoretical interpretation of the law. We undertook a pilot analysis of Bath law using observational material selected to study the relationship $\sigma(M_0)$.

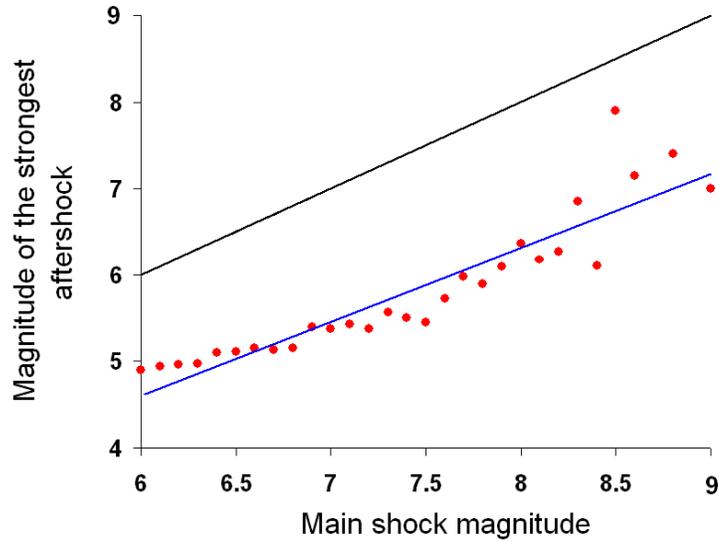

**Fig. 6.** Comparison of the magnitudes of the strongest aftershocks and the magnitudes of the main snocks (points). The blue line approximates the experimental points. The black line is drawn for clarity.

Figure 6 perfectly illustrates the essence of Bath law. The blue line

$$M_{max} = -C + DM \quad (17)$$

approximates the experimental points. Here $M_{max}$ is the magnitude of the strongest aftershock in the interval from 0 to 20 hours, $C = 0.53$, $D = 0.85$, $R^2 = 0.84$. The black line $M_{max} = M_0$ is drawn for clarity. We see a stable gap between the blue and black lines, which probably widens slightly with the rise of $M_0$. To check this possibility, we plotted the dependence of $\Delta M = M_0 - M_{max}$ on $M_0$. However, the coefficient of determination $R^2 = 0.13$ is too small. Thus, we cannot say that $\Delta M$ depends on $M_0$.

On average for all our measurements $\Delta M = 1.6$. This value is about one and a half times higher than what is usually indicated in connection with the Bath law. The discrepancy between our measurements and the literature data is probably due to the fact that our sample of aftershocks is limited to the time interval of 0–20 h after the main shock. In this regard, we would like to draw attention to the incompleteness of the standard wording of the Bath law. Namely, in addition to indicating the value of $\Delta M = 1.1 - 1.2$, one should indicate the probable waiting time for the appearance of the strongest aftershock. It is not enough to say that sooner or later there will be a strong aftershock with a magnitude not exceeding a certain value. It is necessary to answer at least roughly the natural question of when this will happen. There is still additional work to be done in this direction.

## 6. Summary

The rich experience in studying the three classical laws of the physics of earthquakes can be successfully used for the development of the theory and the search for new approaches to the analysis of experimental data. The methodical reception we have applied is simple and consists in the fact that we introduced the concept of the deactivation of earthquake source and reformulated the Omori law, presenting it in the form of an evolution equation. This made it possible to raise the question of the dependence of deactivation factor on the magnitude of the main shock. The main result of the work is that a monotonic decrease in the deactivation factor with an increase in the magnitude of the main shock is theoretically predicted and experimentally observed. We are confident that the discovery of other dependences of this kind will bring us closer to the physical interpretation of the deactivation factor of earthquake source.

*Acknowledgments*. We express our deep gratitude to B.I. Klain and A.D. Zavyalov for numerous discussions of problems in the physics of earthquakes. We are deeply grateful to A.S. Potapov and A.L. Sobisevich for interest in this work. The authors thank the staff U.S. Geological Survey for providing the catalogues of earthquakes. The work was carried out according to the plan of state assignments of IPhE RAS.